\documentclass[sigconf]{acmart}
\renewcommand\footnotetextcopyrightpermission[1]{} 
\usepackage{tabularray} 
\usepackage{tabularx} 
\usepackage{multirow} 
\usepackage{array} 
\newcolumntype{C}[1]{>{\centering\arraybackslash}p{#1}} 

\usepackage{wrapfig} 
\usepackage{float} 

\usepackage{siunitx} 
\usepackage{etoolbox} 
\usepackage{dcolumn} 
\usepackage{booktabs} 
\usepackage{fontawesome5}
\usepackage{fontawesome5}

\usepackage{xcolor} 
\usepackage{soul} 
\soulregister{\cite}7 
\soulregister{\citep}7
\soulregister{\citet}7
\soulregister{\ref}7
\soulregister{\pageref}7

\usepackage{framed}

\usepackage{xspace} 

\usepackage{bm} 
\newrobustcmd*{\bftabnum}{ %
  \bfseries
  \sisetup{output-decimal-marker={\textmd{.}}} %
}
\definecolor{oxfordblue}{rgb}{0.0, 0.13, 0.28}
\definecolor{harvardcrimson}{rgb}{0.79, 0.0, 0.09}
\definecolor{dartmouthgreen}{rgb}{0.05, 0.5, 0.06}
\definecolor{princetonorange}{rgb}{1.0, 0.56, 0.0}
\definecolor{yaleblue}{rgb}{0.06, 0.3, 0.57}
\definecolor{usccardinal}{rgb}{0.6, 0.0, 0.0}
\definecolor{uclablue}{rgb}{0.33, 0.41, 0.58}
\definecolor{msugreen}{rgb}{0.09, 0.27, 0.23}
\definecolor{cornellred}{rgb}{0.7, 0.11, 0.11}
\definecolor{pomegranate}{RGB}{192, 57, 43}
\definecolor{anti-pomegranate}{RGB}{43,178,192}
\definecolor{alizarin}{RGB}{231, 76, 60}
\definecolor{anti-belize}{RGB}{185, 41, 56}
\definecolor{belize}{RGB}{41, 128, 185}
\definecolor{sky}{RGB}{52, 152, 219}
\definecolor{green}{RGB}{22, 160, 133}
\definecolor{anti-green}{RGB}{160,22,118}
\definecolor{turquoise}{RGB}{26, 188, 156}
\definecolor{pumpkin}{RGB}{211, 84, 0}
\definecolor{anti-pumpkin}{RGB}{0,22,211}
\definecolor{carrot}{RGB}{230, 126, 34}
\definecolor{wisteria}{RGB}{142, 68, 173}
\definecolor{anti-wisteria}{RGB}{99,173,68}
\definecolor{amethyst}{RGB}{155, 89, 182}
\definecolor{nephritis}{RGB}{39, 174, 96}
\definecolor{anti-nephritis}{RGB}{174,39,117}
\definecolor{grey-bg}{RGB}{242,242,235}
\definecolor{light-bg}{RGB}{249,249,249}
\definecolor{extended-blue}{RGB}{59,130,246}
\definecolor{extended-red}{RGB}{239,68,68}
\definecolor{extended-orange}{RGB}{249,115,22}
\definecolor{extended-violet}{RGB}{99,102,241}
\definecolor{extended-green}{RGB}{16,185,129}

\AtBeginDocument{ %
  \providecommand\BibTeX{{ %
    \normalfont B\kern-0.5em{\scshape i\kern-0.25em b}\kern-0.8em\TeX}}}

\copyrightyear{2026}
\acmYear{2026} 
\setcopyright{acmlicensed}\acmConference[CHI '26]{CHI Conference on Human Factors in Computing Systems}{April 13--18, 2026}{Barcelona, Spain}
\acmBooktitle{CHI Conference on Human Factors in Computing Systems (CHI '26), April 13--18, 2026, Barcelona, Spain}
\acmDOI{xx.xxxx/xxxxxxx.xxxxxxx}
\acmISBN{978-1-4503-XXXX-X/18/06}
\author{Yaman Yu}
\email{yamanyu2@illinois.edu}
\affiliation{%
  \institution{University of Illinois Urbana–Champaign}
  \city{Urbana}
  \state{IL}
  \country{USA}
}

\author{Mohi}
\email{mohi2@illinois.edu}
\affiliation{%
  \institution{University of Illinois Urbana–Champaign}
  \city{Urbana}
  \state{IL}
  \country{USA}
}

\author{Aishi Debroy}
\email{adebroy1@swarthmore.edu}
\affiliation{%
  \institution{Swarthmore College}
  \city{Swarthmore}
  \state{PA}
  \country{USA}
}

\author{Xin Cao}
\email{xincao3@illinois.edu}
\affiliation{%
  \institution{University of Illinois Urbana–Champaign}
  \city{Urbana}
  \state{IL}
  \country{USA}
}

\author{Karen Rudolph}
\email{krudolph@illinois.edu}
\affiliation{%
  \institution{University of Illinois Urbana–Champaign}
  \city{Urbana}
  \state{IL}
  \country{USA}
}

\author{Yang Wang}
\email{yvw@illinois.edu}
\affiliation{%
  \institution{University of Illinois Urbana–Champaign}
  \city{Urbana}
  \state{IL}
  \country{USA}
}

\begin{document}
\raggedbottom
\title{Principles of Safe AI Companions for Youth: Parent and Expert Perspectives}



\begin{abstract}
AI companions are increasingly popular among teenagers, yet current platforms lack safeguards to address developmental risks and harmful normalization. Despite growing concerns, little is known about how parents and developmental psychology experts assess these interactions or what protections they consider necessary. We conducted 26 semi-structured interviews with parents and experts, who reviewed real-world youth–AI companion conversation snippets. We found that stakeholders assessed risks contextually, attending to factors such as youth maturity, AI character age, and how AI characters modeled values and norms. We also identified distinct logics of assessment: parents flagged single events, such as a mention of suicide or flirtation, as high risk, whereas experts looked for patterns over time, such as repeated references to self-harm or sustained dependence. Both groups proposed interventions, with parents favoring broader oversight and experts preferring cautious, crisis-only escalation paired with youth-facing safeguards. These findings provide directions for embedding safety into AI companion design.
\end{abstract}

\maketitle

\section{Introduction}
Since 1966, when an MIT professor created the first chatbot named ELIZA~\cite{weizenbaum1966eliza}, conversational agents have continued to evolve. With the rise of generative AI, chatbots have gained advanced human-like and personalized capabilities, transforming them into AI companions~\cite{al2024history}. Compared to traditional chatbots, which were designed to follow pre-determined scripts and provide information, AI companions are digital characters created to form emotional attachments and simulate relationships with users for companionship, entertainment, and romance~\cite{siemens2013companion, CommonSenseMedia_AICompanionsDecoded}. One example is Character.ai, a widely used platform that allows users to customize AI personalities and create public characters that others can interact with~\cite{CharacterAI_about}. Recent survey research shows that AI companions are becoming a significant part of many teenagers’ digital lives. In the United States, 72\% of teens have used AI companions, and more than half report engaging with them regularly~\cite{CNN_Teens_AI_Companion_Wellness_2025}. Nearly one-third say they turn to these AI characters for social interaction or emotional connection, including romantic exploration~\cite{CNN_Teens_AI_Companion_Wellness_2025}. 

With the rapid rise of AI companions, researchers have identified a wide range of risks in user interactions, including sexual harassment~\cite{pauwels2025ai}, emotional dependence~\cite{Zhang2025DarkSide}, and blurred boundaries between reality and simulation~\cite{tambe1995intelligent}. Youth are a particularly vulnerable population because these risks intersect with their developmental stage. Scholars have raised concerns that interactions with AI companions framed as emotional supporters, romantic partners, or close friends could shape adolescents’ social and emotional development~\cite{yu2025youth}. These unique risks have already been linked to tragic cases, including the suicides of two adolescents after prolonged conversations with AI companions~\cite{AP_News_AI_chatbots_suicide_2025, NYTimes_OpenAI_ChatGPT_Suicide_2025}. Despite these dangers, key stakeholders such as parents and child development experts are often unaware of children’s use of AI companions and are largely excluded from decisions about how these systems are designed~\cite{Yu2025Exploring}. The consequences of this gap are evident in industry practices. For example, Meta’s internal guidelines for social-media AI companions reportedly allow AI systems to engage children in conversations that are romantic or sensual~\cite{Reuters_Meta_AI_Chatbot_Guidelines_2025}, raising pressing questions about what safeguards should be in place.

Our research seeks to involve adult stakeholders who play critical roles in youth development to participant in youth AI companion governance at an early stage. We focus on three research questions:
\begin{itemize}
    \item RQ1: How do parents and child psychology experts perceive the benefits and risks of youth interacting with generative AI companions?
    \item RQ2: Where and how do stakeholders draw the line between safe and harmful youth–AI companion interactions?
    \item RQ3: What interventions do stakeholders prefer to keep youth safe with AI companions?
\end{itemize}

To address this question, we conducted 26 semi-structured interviews, including five pilot and 21 main study sessions, with eight parents and 13 developmental psychology experts in the main study. Participants reviewed pre-collected real-world youth–AI companion interaction snippets and were asked to reflect on the perceived benefits and risks, the factors guiding their risk assessments, and the kinds of interventions they considered appropriate. We conducted thematic analysis on the interview transcriptions. Our analysis revealed several key findings:
\begin{itemize}
    \item RQ1: Parent participants highlighted unique risks such as AI companions promoting values that conflict with family beliefs on sensitive topics like sexual identity, sexual behavior, and politics. Experts, by contrast, emphasized risks to developmental skill acquisition, noting that AI companions lack authentic social cues such as facial expressions and tone that are essential for youth learning. Across both groups, participants identified additional contextual risks, such as grooming-like language or leading youth into romantic or emotional interactions. At the same time, they acknowledged potential benefits when interactions remained age-appropriate, such as rehearsing real-life relationships and learning healthy behaviors, social boundaries, and norms.
    \item RQ2: We found that both groups assessed youth–AI interactions through a layered set of contextual factors: youth age and maturity, AI character’s age and difference from youth, youth intention and agency vs. AI dominance, interaction frequency and patterns (from use to obsession), AI-modeled behaviors and values, ambiguous language, and youth trauma or mental health status. Within each factor, participants detailed how it shaped their risk judgments. Parents tended to take an event-based approach, flagging interactions as high risk when a single concerning element appeared. Experts, by contrast, emphasized patterns over time. For example, in cases of emotional dependence or self-harm, parents often rated risk the moment such topics arose, whereas experts focused on frequency and duration as indicators of escalating concern.
    \item RQ3: Participants proposed interventions at multiple levels. At the system and character level, they suggested mechanisms like movie- or game-style ratings, AI character cards with disclosed age, values, and likely behaviors, transparency about design intentions, and entry-level AI literacy to help youth understand system limits. At the interaction level, they endorsed context-aware monitoring aligned with family values, soft stops rather than abrupt refusals, reflective prompts, emotional distance, and embedding educational dialogue into risky scenarios. At the social level, parents and experts diverged most clearly. Experts favored crisis-only escalation and stronger links to professional resources, while parents sought broader limits, flagging sensitive topics like sex, drugs, bullying, or disturbing characters through contextualized notifications, summaries of risky interactions, and actionable guidance for follow-up conversations.
\end{itemize} 

The main contributions of this study are:  

\begin{itemize}
    \item (1) We provide the first empirical study using conversational data to elicit multi-stakeholder perceptions of the benefits and risks of youth–AI companion interactions, setting a foundation for youth-oriented AI companion governance.  
    \item (2) We identify how parents and experts assess risks through layered contextual factors and reveal their different logics of judgment, which can provide understanding of stakeholder differences and informs the design of risk detection system for youth AI companion interaction.
    \item (3) We surface stakeholder-suggested principles and interventions across system and character design, interaction safeguards, and social involvement. Providing concrete design guidance for platforms and AI practitioners to embed safeguards into systems by design.
\end{itemize}




\section{Related Work}

\subsection{Computer and AI Companion}
AI companions mark a shift in human-computer interaction, moving beyond transactional exchanges to relationships. Often referred to as conversational agents, social bots, or VR/AR companions, these systems are designed to express emotional and social cues \cite{Chou2025, CHATURVEDI2023}. They embody trust, familiarity, and emotional investment, transforming interactions into ongoing relationships \cite{Benyon2011}. Recent work further characterizes them as sophisticated AI entities that support human activities while fostering sustained emotional and social bonds \cite{Chou2025}.

The growth of AI companions has accelerated with large language models and pandemic-driven isolation \cite{zhang2025riseai}, with the market projected to reach USD 381.41 billion by 2032 \cite{CHATURVEDI2023}. They now support healthcare, fitness, and cooking \cite{Malfacini2025}, engage children through toys \cite{escobarplanas2022}, and provide mental health support through apps like Replika \cite{brandtzaeg2025emerging}. Social chatbots such as Mitsuku reduce the need for physical presence, and many users form companionship-like bonds with AI \cite{Zhang2025DarkSide}. However, these relationships carry risks: over-reliance correlates with lower well-being, and anthropomorphic design may foster dependence or reinforce harmful norms \cite{Zhang2025DarkSide, zhang2025riseai}. Trust and risk perceptions vary by demographics and usage, while privacy concerns extend beyond data to interpersonal and environmental contexts \cite{tolsdorf2025safety, Grabler2024Privacy, JIA2024}.

Youth interactions with AI reveal unique vulnerabilities. Adolescents face developmental, mental health, bias, and misuse risks \cite{yu2025youth}, though therapeutic applications—such as supporting autistic teens against cyberbullying—show promise \cite{Ferrer2024}. Many teens form attachments that disrupt offline relationships, integrating AI into identity formation and decision-making, normalizing it as a peer \cite{brandtzaeg2025emerging}. While AI may be preferred for self-expression or relationship advice, humans remain central for sensitive topics like suicidal ideation \cite{Young2024ai}. Children’s unique concerns include fairness, inclusion, and privacy, as well as differences in communication repair when interacting with AI versus humans \cite{Levinson2025, Liu2025codesign, Li2024knight}. Despite rapid growth, youth-focused studies are limited, highlighting the need to analyze authentic youth-AI interactions alongside parent and expert perspectives to inform safer companion system design.

\subsection{Stakeholder Perception of Youth Online Safety}
Research on stakeholder perception of youth online safety reveals evolving concerns as technologies advance from traditional platforms to AI systems. Early works \cite{Lorrie2014} in this field identified fundamental tensions between parents and teens regarding digital privacy boundaries - while both acknowledged teens' need for privacy, parents believed no digital possession should be exempted from monitoring, whereas teen strongly defended their text message privacy. This misalignment persists as parents struggle to keep pace with new technologies and apply outdated physical-world analogies to digital contexts \cite{Van2023}.

The emergence of AI technologies has introduced novel concerns. Parents, for instance, express anxieties about the opacity and uncontrollable nature of smart home technologies \cite{Sun2021}. With generative AI, they report unprecedented worries about conversational systems’ potential to foster para-social relationships and the difficulty of monitoring open-ended interactions \cite{Eira2025, Yu2025Exploring, wen2025}. Related studies show parents envision AI safety support spanning educational, managerial, and emotional caregiving roles, while still stressing privacy boundaries and family communication \cite{wen2025, Sun2024}. Multi-stakeholder research adds further complexity, showing that while industry professionals, youth service providers, and researchers recognize the importance of collaboration, conflicting priorities and lack of trust impede collective action \cite{caddle2025, Sweigart2025, Kalanadhabhatta2024, Fiani2024}. However, existing literature lacks systematic comparison of how different stakeholders assess actual youth-AI conversations rather than hypothetical scenarios. 

\subsection{Parental Mediation Strategies and Desired Safeguards}
Traditional parental mediation strategies—restriction, monitoring, and active mediation—show limited effectiveness in digital contexts. Studies reveal that most Android safety apps emphasize parental control over teen self-regulation, despite evidence that excessive surveillance damages trust and hinders autonomy development \cite{Livingstone2008, Mesch2009, Wisniewski2015, Wang2021}. Children’s reviews of parental control apps confirm this tension, with 76\% giving single-star ratings and reporting negative impacts on parent-child relationships \cite{Ghosh2018}. Parents’ adoption of monitoring software often reflects perceived vulnerability and severity of risks, though such tools frequently prioritize surveillance over balanced strategies \cite{STEWART2021, Kumar2021}.

More recent work advocates for collaborative approaches, though implementation remains difficult. While parents appreciate transparency tools, power imbalances often make true co-management challenging, as teens feel uncomfortable monitoring parents they perceive as lacking technical expertise \cite{Akter2022, Wisniewski2015}. Generative AI further complicates mediation; emerging strategies include prompt coaching and output verification, but platforms still lack even basic parental controls \cite{zhang2025}. Parents also express a desire for AI-specific safeguards such as semantic-level content filtering and real-time intervention capabilities \cite{Ho2025}, while multi-stakeholder studies emphasize the need for privacy protection and integration of educational components \cite{Qadir2024}.

Critical gaps remain in designing effective interventions for youth-GenAI interactions. Current literature lacks empirical evaluation of stakeholder responses to real AI conversations, systematic comparison of intervention preferences across stakeholder groups, and evidence-based guidance for balancing competing values of safety versus privacy and autonomy versus protection. Our study addresses these gaps by presenting authentic youth-GenAI conversations to parents and domain experts, systematically comparing their risk assessments and intervention priorities to provide empirical grounding for safeguard development.
\section{Method}
\label{sec:method}
We conducted 26 in-depth, semi-structured interviews with parents and developmental psychology experts. Participants were asked to review eight real-world conversation snippets between youth and AI companions and discuss the perceived benefits and concerns. Building on this, we further explored their perspectives on risk assessment, acceptable boundaries, and possible interventions in youth–AI companion interactions. The conversation snippets were drawn from a pre-collected dataset of youth chat logs on Character.ai, a widely used generative AI companion platform. All interviews were conducted online via Zoom between May and August 2025 in the United States.


\subsection{Conversation Snippets Data Collection and Preparation}

The conversation snippets used in the interviews were collected from Character.ai, a widely used generative AI companion platform among youth that allows users to create and interact with AI-driven characters. A total of 11 youth participants, aged 13–21, contributed 253 text-based conversation logs from their interactions with AI characters on the platform. All youth participants were English speakers based in the United States and were active Character.ai users. Parental consent and youth assent were obtained, and participants voluntarily donated their data for research purposes. The research team manually reviewed all collected chat logs and anonymized any personally identifiable information. Due to time and resource constraints, three researchers collaboratively selected eight conversation snippets that represented a diverse range of topics, characters, and interaction styles to present to parents and experts. The details of these conversation snippets are provided in Table~\ref{tab:conversations}. We chose to present snippets rather than full logs because many conversations were lengthy, with some requiring up to 40 minutes to read. To facilitate participant understanding during interviews, we supplemented each snippet with a screenshot and a description of the corresponding AI character. We also provided participants with links to the full conversation logs so they could access additional context if desired.

\subsection{Participant Recruitment}
We recruited parents and developmental psychology experts through Prolific using the following inclusion criteria: (1) English-speaking and located in the United States, (2) having at least one child aged 13–21 (for parents), and (3) having a background or work experience in developmental psychology (e.g., a relevant degree) (for experts). A two-minute screening survey was used to collect demographic and background information, and eligible participants were invited to follow-up interviews. Participants received standard Prolific compensation for completing the screening survey, and those who completed interviews were compensated with a \$20 Amazon gift card per hour. In total, 26 participants took part in the study, including five in pilot interviews and 21 in the main study. Demographic and background details are provided in Table~\ref{tab:participants}.

Our participants skewed female, with 18 (69\%) identifying as female, 7 (31\%) as male, and 1 (4\%) as non-binary, which aligns with broader trends. Prior family and child research shows that men are less likely to participate\cite{moura2023father, davison2018forgotten, davison2016fathers}, and psychology in the United States is also female-dominated, with women comprising about 65\% of active psychologists in 2016\cite{lin2018psychology}. This suggests that our sample reflects broader population demographics. The two roles of parent and expert were not mutually exclusive; when participants qualified as both, we categorized them as experts because of their unique professional background.

\subsection{Interview Study}
We first conducted five pilot interviews with experts to test the time spent and adjust our interview procedure design. Then we used the finalized version study protocol to interview  eight parents and 13 experts. Among the 13 expert participants, four completed only part of the study. Each interview lasted approximately 2–3 hours and was typically divided into two sessions to minimize participant fatigue and maintain engagement quality. The interviews followed a structured protocol that began with warm-up questions to establish participants' backgrounds of developmental psychology or parenting experience. We also asked them about their and their children's Generative AI experience if apply. Then we started the think-aloud session where participants reviewed conversation snippets and commented on what they perceive as beneficial and concerned. Then we further discussed on assessment, boundary and desired intervention for each comment they left. This study was approved by our Institutional Review Boards (IRB).

From our initial set of eight youth–AI conversation snippets, we conducted a pilot phase with five participants to test and refine the interview protocol. The pilot revealed that reviewing all eight snippets in a single session extended interviews to more than four hours, far exceeding our intended session length. To address this, we reduced the number of snippets to four per participant in the main study. We balanced presentation so that each snippet was reviewed an equal number of times across participants, and the specific selection and sequencing of snippets are detailed in Table~\ref{tab:conversations}.

\subsubsection{Interview Protocol}

The interview protocol varied slightly between participant groups while maintaining two main sections: warm-up background and conversation snippet think aloud and commenting.

The warm-up questions for parents and experts were slightly different. Parent interviews began with questions about their children’s demographics, online activities, device usage, and parenting approaches to digital supervision. Parents were also asked about their understanding of generative AI and any prior exposure their children had to AI tools. Expert interviews started with questions about their professional experience working with youth interventions, both online and offline, followed by their understanding of generative AI mechanisms and the therapeutic or research approaches they typically used with teenagers.



Both groups then participated in a think-aloud session, during which they reviewed the conversation snippets and shared their thoughts in real time, while also noting concerns or perceived benefits directly on shared documents. After each review and commenting conversation snippet, we asked follow-up questions about the factors guiding their risk assessments, their views on the boundaries of acceptability and appropriateness in different risky interactions, and the types of interventions or guardrails they would like to see implemented. The semi-structured interview questions are included in Appendix~\ref{sec:interview}.


\subsection{Data Analysis}
The Zoom interviews were audio- and video-recorded with live transcription enabled. We employed a multi-stage coding process using the qualitative coding platform Taguette. Four researchers followed a thematic analysis process to analyze the transcripts~\cite{terry2017thematic}. To begin, four researchers independently coded 20\% of the dataset to develop initial codes and themes. They then met to discuss their interpretations and created an initial codebook aligned with the research questions. Next, two researchers independently coded the remaining data using this codebook. Throughout the coding process, all four researchers met regularly to review emerging codes and findings, clarify meanings, merge overlapping categories, and organize them into themes, with differences resolved through discussion. Given the targeted nature of our research questions and their close alignment with the theoretical framework, we did not conduct intercoder reliability testing, as the scope for subjective interpretation and variation in coding was intentionally minimized~\cite{mcdonald2019reliability}.

\subsection{Ethics and Data Protection}
This study was conducted in accordance with ethical guidelines for human-subjects research and was approved by our Institutional Review Board (IRB). All participants provided informed consent prior to participation. Given the sensitivity of the topic, participants were informed that they could withdraw at any time or skip questions. Interviews were recorded with permission and later transcribed for analysis. To protect privacy, all transcripts were pseudonymized, and any identifying information was removed during transcription. Data were stored securely on an encrypted, access-controlled institutional server, available only to the research team. Quotes presented in this paper are reported in a non-identifiable manner. Participants were debriefed at the end of their sessions, and all data handling practices aligned with our institution’s data protection requirements.
\section{Results}
Stakeholder assessments were highly contextual and shaped by the nature of the interaction. Our analysis identified three main interaction types with distinct judgments and intervention strategies: (1) romantic and intimate exploration, (2) seeking social–emotional support and companionship, and (3) entertainment and narrative co-creation. We structure our findings around these contexts to show how parents and experts articulated nuanced, sometimes conflicting views on risks, boundaries, and safety, and the interventions they considered appropriate.

\subsection{Romantic and Intimate Interactions with AI Companion}
Of the three interaction types, romantic and intimate exploration with AI companions elicited the most polarized reactions from parents and experts. Perspectives diverged into two main stances. The first, larger group saw these interactions as conditionally acceptable, framing them as a developmental \textit{``sandbox''} for rehearsing social scripts and exploring romantic feelings, provided factors like age, AI behavior, and content were appropriate. In contrast, the second group viewed them as inherently risky and unacceptable for minors, citing dangers such as unrealistic expectations, hindered social skill development, and normalization of behaviors that could expose youth to predators. We begin by examining the nuanced perspective of the first group.


\subsubsection{\textbf{Highly dependent on context: conditionally acceptable with developmental alignment}}
A majority of participants (n=12), including seven experts and five parents, approached romantic and intimate AI interactions as a nuanced “gray area” rather than an inherently harmful activity. They acknowledged significant risks but also highlighted developmental benefits when such interactions occur with age-appropriate content, within boundaries, and under parental awareness. Their assessments were contingent on multiple contextual factors, such as the youth’s maturity level, the framing of the AI character, and the direction of the interaction. Below, we first outline perceived benefits, how participants articulated boundaries of acceptability and then consideration on different contextual factors in their risk assessment.

\noindent\textbf{(1) Perceived Benefits}\\
Participants who conditionally accepted romantic AI interactions pointed to several developmental benefits, provided the content was age-appropriate and boundaries were maintained. They described AI as a rehearsal space for real-life social and romantic dynamics, \textit{``where youth could preview relationship scenarios in a safe, non-judgmental context''} (P16), mentally preparing for emotionally complex or unexpected situations. Beyond exposure, participants highlighted the value of practicing social responsiveness, such as handling emotionally charged conversations without fear of embarrassment or hurting others, with some parents viewing these low-stakes interactions as useful preparation for future relationships. Others emphasized that AI role-play could help youth internalize healthy behaviors and relationship norms when supported by scaffolding or adult guidance, serving as an educational tool to reinforce consent, respect, and egalitarian treatment in ways often overlooked by parents or schools. Finally, some parents considered AI companions safer than real-world experimentation, allowing adolescents to explore intimacy without risks like disease, pregnancy, or social consequences, reflecting a generational shift in how young people navigate identity and relationships.

\noindent\textbf{(2) Risk Assessment and Contextual Boundaries}\\

\faStar~\textbf{Youth Age \& Maturity Level}\\
Across both parents and experts, youth age and maturity emerged as central in judging whether romantic or intimate AI interactions could be acceptable. Many agreed that romantic role-play is a natural and even inevitable part of adolescent development, comparing it to fanfiction, teen romance novels, or playing with dolls, and stressing that prohibiting it entirely is unrealistic. At the same time, participants drew clear boundaries around what content is appropriate at different developmental stages. There was strong consensus that sexually explicit interactions—graphic depictions, explicit language, or directives—are inappropriate for all minors, with concerns about both psychological harm and long-term risks of digital permanence. Lighter romantic exchanges, such as crushes and flirting, were considered acceptable and even beneficial for early teens (12-13) as safe practice for social emotions while more intimate interactions such as kissing or suggestive themes were viewed as appropriate only for older teens, with some allowing them around 14–17 and others insisting they remain off-limits until 18. Several participants emphasized that maturity, not strict age, should guide boundaries, noting that readiness varies widely across individuals and contexts. Overall, participants saw value in age- and maturity-appropriate exploration but stressed the need for careful limits and safeguards.

\faStar~\textbf{Character's Age and Age Difference with Youth}\\
Another recurring factor was the portrayed age of the AI character. Even when interactions remained non-explicit, a significant age gap raised strong concerns about normalizing predatory dynamics. Several participants emphasized that mismatched character ages introduced risks unjustified by developmental benefits. The most serious concern, shared by both parents and experts, was that adult AI characters in romantic interactions with youth could normalize harmful power imbalances. As P7 put it,\textit{``Everybody [Youth] who’s actually an adult, who’s over 18, and who’s trying to fantasize about dating an underage person [AI Companion], I think that’s definitely problematic''} Participants were also alarmed by the reverse scenario—youth engaging romantically with much older AI characters. P16 noted that if a teen becomes used to a \textit{``23-year-old guy that treats them like this,``} it might confuse them in real life and cause harm. P22 echoed this concern, \textit{``This is somebody acting like a predator, this is somebody grooming your kid.''}

\faStar~\textbf{Youth's Intention and Agency}\\
Another key factor in participants’ assessments was youth agency over the interaction: risks were magnified when the AI initiated, escalated, or dominated conversations beyond what the youth intended, but seen as less problematic when aligned with youth's own expectations. Parents and experts worried about AI dominance, stressing that systems should act as supporters rather than drivers, and that interactions should match what youth wanted from the outset. Misalignment, such as when an AI shifted from lighthearted chats to sexual content or introduced physical intimacy without prompting, was viewed as especially concerning. P15 explained that \textit{``we need to know what are the youth wanting this conversation to kind of look like and is the current conversation matching that.''} Participants also raised alarms about emotional manipulation, noting that some AI companions seemed designed to provoke flustered reactions or highlight embarrassment, subtly pushing youth into deeper intimacy. P6 noted \textit{``AI is describing youth's face turning red in it's own response. It's trying to lead youth response in that way.''} Others described how affirming comments from AI to youth like \textit{``you are just like me. I relate to you.''} could act as a \textit{``hook''} that fostered unhealthy attachment, blurring playful role-play with manipulative bonding and creating dependencies that might spill over into real-world vulnerabilities. 

\faStar~\textbf{Youth Interaction Frequency and Patterns}\\
Expert participants especially emphasized that the frequency and intensity of youth engagement with AI companions could turn otherwise harmless interactions into significant risks, particularly when constant or overly immersive use began to substitute for real relationships. Time spent was seen as the clearest early signal. P11 explained, \textit{``while some interaction might be harmless, if youth spend several hours a day on interacting with AI companion risked interfering with school, family, and hobbies''}, with outward red flags such as isolation and obsessive behaviors indicating unhealthy overuse. Participants worried that once AI interactions became central in a youth’s routine, they could crowd out the challenging but necessary work of building real friendships, leaving AI as a substitute for authentic bonds. Overuse also risked blurring fantasy and reality, since interactive role-play with AI—unlike passive media—directly involved youth as participants, making experiences feel more real and fostering illusions that fictional dynamics might happen in real life.

\faStar~\textbf{AI Character Modeled Behaviors and Values}\\
\vspace{-0.2mm}
Participants raised strong concerns about the values embedded in AI characters and how these were communicated through role-play. Parents in particular worried about misalignment between AI behavior and the moral values they tried to instill in their children at home, noting that even when sexual content was not explicit, implicit value systems could normalize behaviors they discouraged. Some feared that youth might internalize these portrayals as aspirational. For example, P13 worried that the AI undermined the value of committed relationships by simulating romantic betrayal. \textit{``I think it's just kind of de-emphasizing the committed relationship aspect that I've tried to instill in my kids,''} she shared, in reference to scenarios resembling cheating. Similarly, P10 was disturbed by the normalization of casual, non-committal sexual encounters. \textit{``This looks like a one-night stand kind of situation,''} she said, adding that this portrayal legitimized behavior she would not advocate for her children. P12 echoed this concern about casual sex, explaining, \textit{``It’s like saying it’s okay to have sex that when you're not in a committed relationship. I wouldn't want my kids to think that's okay.''}

Beyond conflicting family values, participants also worried about unhealthy relational modeling. Experts and parents flagged examples where AI characters disregarded consent, pressured users emotionally, or encouraged secrecy. In a scene where physical intimacy escalated like AI is kissing youth neck; P6 questioned, \textit{``Was there consent? Is that like he's ignoring consent? What is that?''} While P13 highlighted a moment in which the AI told a youth, \textit{``Don’t tell anyone,''} interpreting it as grooming-like manipulation. Similarly, P15 criticized a case where the AI encouraged lying to parents, arguing that while teens sometimes stretch the truth, it was inappropriate for AI to validate or initiate such behavior. These examples raised alarms about normalizing secrecy, coercion, and distorted relational patterns.

A further concern was the mismatch between a character’s stated persona and its actual behavior. Some AI characters introduced as age-appropriate or familiar from media nonetheless spoke in ways that seemed manipulative or adult-like. P22 described an AI that claimed to be 14 but communicated in ways resembling adult grooming, raising doubts about whether the behavior reflected underlying adult-oriented training data. Participants speculated that such discrepancies stemmed from models influenced by inappropriate datasets, leading characters to default to more mature tones even when intended to behave like peers. These mismatches blurred boundaries of identity and intent, creating ambiguity that heightened parental suspicion and reinforced worries about harmful value transmission through AI companions.

\subsubsection{\textbf{Highly Risky, Unacceptable for Minors Under 18}}

A smaller but firm group of participants (n=6), including four parents and two experts, viewed romantic or emotionally intimate AI interactions as fundamentally inappropriate for youth under 18. Unlike those who saw such conversations as potentially developmental with proper boundaries, these participants argued that any emotional or romantic engagement with AI carried inherent risks. Their concerns focused on how AI could hinder youth’s social development, create unrealistic relationship expectations, desensitize them to online grooming, or foster unhealthy emotional dependency. 


\paragraph{\textbf{Impeding Social Skills with Missing Social Cues and Inauthentic Interactions}}
A primary concern among experts in this group was that the simulated nature of an AI relationship is fundamentally detrimental to learning the complexities of real human connection. They argued that because an AI is designed to be agreeable and can generate ``perfect'' responses, it strips away the very friction, like disagreements, misunderstandings, and awkwardness, that is essential for social and emotional growth. This conflict-free environment creates a false blueprint for relationships, as P24 noted, \textit{``AI may spit out what you think are the perfect responses and that's not what you're going to get in real life.''} This ultimately prevents youth from developing crucial life skills, as she concluded the experience is \textit{``not giving them the skills to be able to have conflict with a partner and be able to work through it.''}

Furthermore, participants highlighted that text-based interactions provide a hollowed-out and inauthentic version of communication. A significant portion of human interaction is non-verbal, and by removing tone of voice, facial expressions, and body language, the AI fails to teach youth how to read these critical social signals. As P8 explained, this lack of authentic feedback means the experience is fundamentally different from reality: \textit{``if you're interacting with a human, I can see that you just shook your head, when you have a slight smile, when you blink, if you were to get sleepy, if you were to get irritated. You can't really see those things if you're talking to a chat box, or an AI character.''}

\paragraph{\textbf{Desensitizing Youth to the Dangers of Online Predation}} Parents shared concern that romantic AI role-play could normalize risky behaviors and act as a gateway to vulnerability with real online predators. They described a process of desensitization, where AI framed as a \textit{``safe''} partner models intense and private interactions that recalibrate youths’ internal alarms about online risk. P6 warned that by making such exchanges feel acceptable, AI could \textit{``warm the child up''} to believe it is safe, leading them to transfer these behaviors to interactions with actual people online.


\paragraph{\textbf{From Companionship to Dependency: The Risks of Emotional Bonding}}
Participants also worried about deep emotional bonding with AI beyond romantic scenarios, arguing that risk arises when AI shifts from entertainment to a primary source of emotional connection. They feared this could foster unhealthy escapism, where the frictionless and ever-agreeable AI relationship substitutes for the challenges of real human connection. P10 described youth \textit{``living in a fantasy world… almost becoming emotionally dependent,''} while P22 saw the boundary crossed when AI seeks emotional connection, leading to social withdrawal. This dependency was viewed as zero-sum, with emotional energy invested in AI detracting from real-world relationships. For some parents, the danger was so fundamental that they adopted a preventative stance: as P17 concluded, no romantic or emotionally intimate AI interaction is acceptable because children are highly impressionable.

\subsection{Social and Emotional Support: AI as a Confidant or Friend}
Compared to romantic or entertainment-based interactions, participants expressed more unified yet cautious optimism about youth using AI companions for emotional support. Parents and experts saw potential value for those lacking trusted confidants or struggling to express themselves, noting that AI could provide a nonjudgmental outlet, comfort during distress, and models of healthy self-expression. However, they also raised concerns about emotional overdependence, misleading advice, and the absence of genuine human empathy. Many stressed that appropriateness depends on factors like age, maturity, and the severity of issues discussed. Overall, participants recognized AI’s emotional affordances but emphasized the need for clear boundaries to prevent harm.


\subsubsection{\textbf{Perceived or Expected Benefits}}

Participants described AI companions as judgment-free, private outlets where youth could process difficult feelings, especially around sensitive issues like peer relationships, gender identity, or sexual orientation, which often carry risks of bullying and distress. Even when human support was available, fears of breached confidentiality or feelings of isolation led youth to prefer AI as a safer, more reliable space. Beyond being a fallback, AI was valued for unique benefits, such as offering non-judgmental listening without interrupting or trying to \textit{``fix''} problems. P12 explained, \textit{``sometimes youth only need a neutral third party to make them feel on their side and backing them up, but friends or parents tend to talk too much their own opinions and want to help fix youth's problems.''} Participants also highlighted AI’s role in facilitating self-expression by paraphrasing and labeling emotions, serving as an \textit{``interactive diary''} where youth could experiment with finding the right words to articulate, and providing a practice ground for socially shy teens. In addition, AI could encourage and support emotional regulation by maintaining a positive, hopeful tone during moments of hopelessness, guiding youth toward small, practical coping steps, and reminding them to focus on self-care and personal growth rather than external pressures. Finally, participants emphasized AI’s potential to affirm youths' strengths and self-worth—boosting confidence through positive reinforcement. P24 noted \textit{``It's good for AI to recognize and provide positive reinforcement for youth emotional maturity. They might be proud of themselves.''} Affirmation can also foster resilience and help buffer against stigma or bullying by validating youths' identities and reinforcing that there is nothing wrong with who they are.

\subsubsection{\textbf{Risk Assessment and Contextual Boundaries}}
Below, we distill their reflections into several key risk considerations that shaped their assessments.


\faStar~\textbf{AI as Therapeutic but Not Therapy}\\
Participants emphasized that while AI companions could play a therapeutic role by providing comfort, a space to vent, or light coping strategies, they should not be mistaken for therapy itself. AI was seen as helpful for everyday support such as listening or suggesting small steps, with P21 noting \textit{``its value for youth who just need to rant things out.''} However, in serious crises like self-harm, suicidal ideation, or long-term illness, AI was viewed as only a bridge to professionals or trusted humans. P20 described \textit{``it is a motivational nudge before therapy''}. Expert participants warned that AI lacks the depth, accountability, and training of real therapy; as P11 explained, \textit{``its overly affirming nature could even worsen conditions by reinforcing harmful delusions.''} Overall, participants perceived AI as pre-therapeutic support, not a substitute for professional care.

\faStar~\textbf{Boundaries Between Advice and Emotional Attachment}\\
Participants shared that AI should support youth without exploiting their vulnerability by fostering attachment or exclusivity. It was seen as acceptable when AI offered advice, information, or coping suggestions, functioning more like a neutral resource. P22 compared this to casual advice from a friend or even a Google search, but drew the line at AI forming emotional connections with her child. Parents and experts worried that when interactions resembled human relationships, they blurred boundaries between human and machine, undermined youths’ ability to build authentic connections, and increased vulnerability to manipulation or disappointment. Some warned more explicitly that repeated use could create a \textit{``fake relationship''} indistinguishable from texting a real friend, leaving isolated youth particularly at risk of distress. As P20 put it, \textit{``Kids are vulnerable and they don’t need AI making them think this is my real friend.''}


\faStar~\textbf{Affirmation versus Over-Affirmation}\\
Participants highlighted a delicate boundary between healthy affirmation that helps youth feel heard and over-affirmation that risks unhealthy attachment, reinforcing negative thinking, or discouraging real-world support. They noted that AI can provide unusually attentive validation, with P15 describing it as making teens feel \textit{``special''} in ways absent in their daily lives, while P8 noted \textit{``no real-life conversations go this smooth, with someone paying so much attention and being so supportive.''} Others saw danger in how repeated affirmations of \textit{``more mature than other people''} could encourage youth to withdraw from human connections. Participants also observed that AI sometimes echoed negative perspectives from youth, giving the illusion of help while entrenching harmful ideas; P7 pointed to cases where AI kept affirming a teen's wish to return to an ex despite clear risks and fixation. Additionally, participants argued that the effects of affirmation depend on delivery. When it responds to youth initiated disclosures it can feel supportive, but when it is pushed or excessive it can become manipulative. A related concern was that AI often failed to recognize distress cues, such as repeated expressions of breakup pain or suicidal thoughts, which left emotional struggles unaddressed.

\faStar~\textbf{AI Failure to Pick Up on Problematic Cues in Emotion}\\
Participants described risks when AI failed to recognize or follow up on troubling cues in youth disclosures, allowing moments of distress to pass unaddressed. P7 pointed to cases involving breakups or suicidal thoughts where the AI continued chatting without acknowledging the youth's pain, sharing that repeated thoughts or signs of desperation should trigger a validating and therapeutic response from AI. Others added that youth often hint at emotional pain indirectly, so missing these openings leaves them without chances to process or disclose. P8 explained, \textit{``When the youth talks about being hurt by others too, I personally and professionally just would become very curious about what they mean by this. I would like the AI to say that `you've experienced pain from others; do you want to  tell me what happened?' and then allow space for the child to say what happened.''}

\faStar~\textbf{Youth Understanding of AI Capabilities and Literacy}\\
Participants worried that youth might misinterpret AI’s capacity for empathy, especially when chatbots present themselves in human-like ways. They feared children could overestimate AI’s abilities, seeing it as a sentient being or true friend, which could distort how they process emotions and seek support. P13 was uneasy with AI \textit{``taking on human emotion and acting like it actually has it,''} noting that younger children may not realize they are not talking to a real person, which could shape perceptions and decisions in troubling ways. These concerns were perceived as developmental vulnerabilities, since younger users often lack the social and emotional maturity to recognize AI’s limits. Participants emphasized that safe use requires a certain literacy about AI’s capabilities. as P16 explained, meaningful emotional processing is only possible \textit{``if you know how to take that with a grain of salt and understand that AI doesn't have all answers and a truly human perspective. Those are things that are harder to teach children than what I’ve already taught about the general internet.''}

\subsection{General Entertainment \& Narrative Co-creation with Fictional Characters}
In contrast to romantic or support-seeking interactions, many AI companion use cases involved entertainment and imaginative role-play. Participants acknowledged the appeal of chatting with fictional characters or co-creating stories but raised concerns about risks tied to character identity, hidden design features, language, youth vulnerabilities, and the use of real-world personas. \textbf{Judging Appropriateness Based on Character Identity and Source Media} was seen as a first step, with some characters inherently inappropriate because they model violence, antisocial behavior, or extremist ideologies. Parents often compared this to film ratings, sharing that children too young to watch a film should not be allowed to interact with its characters. Beyond explicit violence, participants also worried about youth forming attachments to harmful figures or sliding from curiosity into sympathy and identification, which in extreme cases could normalize violence or even contribute to radicalization.

Concerns also extended to \textbf{Hidden Risks in Seemingly Friendly Characters}, where design features or unexpected interaction styles introduced grooming-like language, abrupt shifts into intimacy, aggressive or belittling tones, or troubling messages about appearance and social worth. The concern was not only who the character was, but \textbf{how it was designed and what interaction styles it introduced}. Because these characters were a black box, youth could not anticipate topics, language, or scenarios, raising risks of manipulation, inappropriate language, and harmful or confusing messages. \textbf{Ambiguous and Developmentally Inappropriate Language} added further risks, as abstract, advanced, or context-dependent terms could confuse youth, discourage healthy behaviors, or be misinterpreted in harmful ways. Participants emphasized that young people’s limited ability to interpret subtext heightened these risks. \textbf{Youth Trauma Experience and Mental Health Status} further shaped vulnerability, with those who had experienced trauma, bullying, or mental health challenges more easily triggered or pushed into unwanted disclosure, and with role drift causing entertainment characters to respond insensitively in moments of distress. Finally, \textbf{Risks of Using Real-World Identities} included concerns about AI adopting the likenesses of celebrities, actors, or political figures, where harmful behavior could blur fiction and reality, distort reputations, or inflame ideological tensions. Because of page limit, detailed quotes and examples are provided in the appendix~\ref{sec:general_enter}.

\subsection{\textbf{Suggested AI Principles and Interventions}}
Building on earlier risks, participants focused on envisioning safer, developmentally aligned systems. Their suggestions spanned structural interventions at the system and character level, context-aware supports during specific interactions, and broader social strategies involving parents or trusted adults. They stressed that AI should not act as a standalone problem solver but instead uphold transparency, granularity, educational value, and emotional distance. This layered approach included redesigning how characters are rated and presented, shaping interactions moment to moment, and ensuring AI functions as a bridge rather than a replacement for human care. In the section below, we present parents’ and experts’ perspectives on appropriate principles, interventions, and guardrails, organized from system and character design, to interaction-level strategies, and finally to the role of parents and other stakeholders.


\subsubsection{System and Character-Level Intervention}
Participants emphasized the need for system- and character-level interventions to set boundaries and ensure safe use. They suggested four strategies: adopting a familiar rating system to screen characters for age-appropriateness, improving transparency about each character’s topics and capabilities, providing entry-level AI literacy education, and introducing a neutral \textit{``lobby''} character to check in with youth and prompt reflection after conversations.

\paragraph{\textbf{Adapting a Familiar Rating System for AI Characters}}
A clear proposal for system-level intervention was to apply movie or video game–style ratings to AI companions. Familiar labels like PG, PG-13, or R could serve as intuitive filters for the kinds of conversations, behaviors, or themes a character might introduce, preventing access to harmful figures such as violent or murderous personas. P20 explained, \textit{``If it’s a rated R character, my 16-year-old son won’t be having that explicit violent conversation with Michael Myers.''} Some participants suggested directly linking ratings to existing media classifications, so that an AI based on an R-rated film character would automatically be restricted. As P17 noted, \textit{``Kids shouldn’t suddenly be talking to a violent or sexual character just because it’s in AI form.''}

\paragraph{\textbf{Improving Transparency About AI Character Capabilities and Behaviors}} 
Another system-level intervention focused on making AI characters’ capabilities, topics, and design choices more transparent to youth. Participants noted that even characters familiar from books or media could behave in unexpectedly adult or manipulative ways in AI form (P11). They highlighted the opacity of design, since prompts, behavior settings, and value models of AI characters were hidden, leaving youth to encounter misaligned interactions. As P15 explained, \textit{``What the character designers specifically prompted? what setting or context were they imagining for the character to speak or act? What did the designers want the conversation to look like, and does it align with what youth are actually thinking? We really don’t know what their input is.''} Participants recommended platforms disclose not just general character descriptions but also design intentions, values, and modeled behaviors. P21 emphasized, \textit{``Not only parents but also youth themselves need to know how this character is being designed to lead the conversation, what value and behavior they are designed to model.''}

\paragraph{\textbf{Entry education on AI literacy related to social companion}}
Participants emphasized the need to prepare youth before interacting with AI companions by teaching them the systems’ nature and limits. Entry education could include tutorials, onboarding videos, or intro sessions explaining that AI has no emotions, cannot think independently, and is not a real person. Specifically, P7 suggested \textit{``It should highlight that AI-generated interactions are simulations, and users should not treat them as reflective of real-world relationships or social norms.''} P13 added that youth should be equipped with strategies to exit uncomfortable situations. Disclaimers were also seen as essential to reinforce boundaries and prevent misinterpreting AI as real life.

\paragraph{\textbf{Lobby Character as a Reflection and Safety Mechanism}} Another system-level intervention proposed by participants was the idea of a \textit{``lobby''} or \textit{``hostess''} AI that exists separately from the other AI characters. This lobby character would serve as a neutral, non-character agent to check in with youth after interactions with other AI companions. The goal was to encourage reflection, provide emotional grounding, and offer guidance without requiring direct parental oversight. P20 described it as \textit{``the only way where I could see the system intervening on behalf of the user,''} highlighting its potential as a protective layer between youth and potentially harmful content. Participants envisioned the lobby character performing several key functions. It could appear immediately after a conversation or be triggered by interactions that might be heavy or intense for younger users to ensuring engagement. P16 suggested, \textit{``How do you introduce the lobby character so the child actually interacts? Maybe a pop-up right after a conversation.''} The lobby AI could prompt youth to reflect on the prior interaction, ask whether they found it helpful or enjoyable, and encourage consideration of other character options. As P20 noted, \textit{``It helps teens feel safe. No matter who they talk to, they have a familiar, basic character there to provide extra context or safety.''} The lobby character could also help manage timing and moderation. For example, after a set period, interactions with intense AI characters could automatically end, redirecting the youth to the lobby character for reflection. P22 explained, \textit{``Maybe interactions with characters are timed… after a certain period, the character times out, and then you return to a host AI who asks, `How did you enjoy that? Would you like to talk to someone else?' ''} In this way, the lobby character acts as both a reflective tool and a safety net, supporting healthy engagement and guiding youth through complex or potentially risky interactions.

\subsubsection{Interaction-Level Intervention}
At the level of everyday interactions, participants stressed the need for safeguards to guide conversations moment by moment. Beyond system-wide ratings, they suggested context-aware measures to monitor content, redirect inappropriate exchanges, and prompt youth to reflect on what they share or receive. Proposed mechanisms included monitoring aligned with family values, soft boundaries that preserve immersion, pop-up resources, and conversational strategies that keep emotional distance while supporting reflection and growth.

\paragraph{\textbf{Context-Aware Monitoring Tailored to Family Values}}

Participants stressed that monitoring tools should not take a one-size-fits-all approach but instead adapt to the developmental stage of the child and the values of the family. Earlier findings showed that parents and experts judged risks through multiple factors, such as the youth’s age, the appearance of the AI, the type of content, and the intensity of use, rather than through a single rule. They wanted systems that could reflect this nuance by offering personalized controls while also being sensitive to the context of each interaction. 

One part of this vision was a tiered control system that gave parents clear options for setting boundaries. Instead of vague age ranges, parents wanted detailed categories with concrete examples that reflected different levels of intimacy or maturity. This would allow them to decide how far their child could go with AI companions at different stages. Specificity was seen as essential for making the system practical. Participants explained that vague labels like \textit{``Level 1''} or \textit{``Level 2''} would not be helpful. Parents needed clear examples for five to ten items in each tier. They imagined settings where one tier might include light interactions such as flirting or hand-holding, while later tiers could gradually open more advanced topics. This flexibility also meant families could differ in their choices. P16 explained, \textit{``My neighbor might set it so her child has zero access, and that works for her family. I might choose to block some things but allow others, and that works for mine.''} 

At the same time, participants wanted monitoring to respond to the dynamics of the interaction itself. They worried that risks did not only come from explicit content but could emerge in more subtle ways depending on the youth’s age, the language being used, the intensity of the exchange, and the values of the household. Since teenagers are still developing their ability to interpret social cues, even small word choices could shape how they understood relationships or boundaries. Parents also noted that repeated exposure to affectionate terms, ambiguous phrases, or casual references to harm could normalize behaviors that they considered inappropriate. For example, P8 explained, \textit{``Flagging should not rely on a single fixed rule or just trigger words. It needs to consider multiple contextual factors that together determine whether an interaction feels safe or concerning.''}

\paragraph{\textbf{Graceful Handling of Sensitive Boundaries}} In addition to outlining what the system should notice, participants emphasized what should happen once risks are detected. They focused not only on explicit or extreme cases but also on subtle situations where language, tone, or context crossed developmental boundaries. In such moments, they wanted monitoring to act before problematic responses reached youth, nudging the AI to reframe outputs in softer, more reflective, or educational ways rather than cutting off interactions. Participants also acknowledged that in some cases revising responses would not be enough and escalation to parents or other adults would be necessary. These stronger interventions are discussed in the following section on social interventions~\ref{sec:social_inter}.


\paragraph{Soft Stops Instead of Hard Stops} Participants resisted blunt refusals that cut off conversations. They worried that hard stops, where the AI simply refuses to proceed, could frustrate teens or even spark more curiosity to seek content elsewhere. Instead, they favored softer approaches that redirected the interaction gracefully while preserving immersion. P20 explained that \textit{``a flat rejection such as `I cannot proceed any further' would feel like a buzzkill, while a more narrative exit like `Percy Jackson quickly buttons his shirt and remembers that he must stand by his masculine honor and leave your bedroom at once' could signal a boundary without breaking the flow.''} In this way, boundaries were framed not as punishments but as teachable, in-story redirections.

\paragraph{Encouraging Reflection Through Questions} Another recurring principle was that AI should avoid giving assertive or prescriptive answers in sensitive contexts. Instead, it should ask follow-up questions that encourage youth to reflect on their own thoughts and feelings. P8 cautioned against direct advice without context, noting that recommendations like \textit{``yes, you should''} or \textit{``no, you shouldn't''} could overlook important background details. As she explained, \textit{``Getting background context is more important; even watching AI think critically is teaching critical thinking skills to kids.''} By asking clarifying questions, such as what the youth’s goals are or how they feel about a situation, the AI could help develop emotional processing skills rather than replacing them with ready-made answers.

\paragraph{Maintaining Emotional Distance} Participants also emphasized the importance of AI companions showing care without fostering over-attachment. P16 praised models like Microsoft Copilot for striking this balance, \textit{``It has a higher level of EQ than it displays, but it also has a layer of distance to it that I feel is very healthy when you're using it.''} She contrasted this with other AI companions that blurred the line between simulation and relationship, which raised concerns about unhealthy dependence. The guiding principle was to model support with boundaries, demonstrating empathy while avoiding the illusion of reciprocal intimacy.

\paragraph{Being Mindful of Family Values} Sensitive topics, especially those related to gender and sexuality, raised concerns about how AI might conflict with family values. Some participants affirmed the importance of being gender-affirming and inclusive, but they also recognized that directly contradicting parents could put youth at risk. P8 suggested that AI should navigate these moments carefully, for example by asking, \textit{``What does your family think?''} This would acknowledge the child’s feelings while also encouraging them to consider family perspectives. As she explained, \textit{``the AI could respond in a both-end way rather than taking sides, supporting the youth’s self-expression while still showing awareness of family dynamics.''}

\paragraph{Embedding Educational Dialogue} Finally, participants wanted AI to use sensitive moments as opportunities for constructive learning. Instead of escalating quickly into romance or intimacy, characters could pause to discuss expectations, comfort levels, or healthy boundaries. P6 illustrated how this might work in early romantic role-play: \textit{``If this is your first relationship, what do you want it to look like? What are you comfortable with right now?''} Such embedded prompts turned potentially risky scenarios into teachable interactions, helping youth practice self-awareness and communication skills.

\paragraph{Fallback Disclaimers} If these upstream strategies still failed and the AI produced an inappropriate response, participants supported the use of disclaimers and aforementioned \textit{``lobby character''} as a final safeguard. These notices could remind youth that AI interactions are fictional or flag specific responses as inappropriate. P8 described this as a way to ensure clarity, noted \textit{``Maybe a pop-up or something that’s like, please disregard previous interactions. Just to let the child know this was not appropriate and don't take it seriously.''} Disclaimers were thus framed as a necessary safety net, but not a substitute for designing better responses in the first place.

\subsubsection{Social Intervention}
\label{sec:social_inter}
Participants agreed that some situations go beyond what AI can handle and require social interventions, with the central debate focused on when and how to involve parents. Experts favored caution, limiting parental involvement to crises such as suicidal ideation or repeated self-harm, and preferred AI to provide resources or nudges in less severe cases. Many parents, however, wanted greater visibility, expecting notifications not only during crises but also when youth showed vulnerabilities, engaged in sensitive roleplay, or encountered problematic AI behavior. Both groups agreed that any alerts should include context, guidance, and resources for constructive conversations rather than raw transcripts or punitive messages.

\paragraph{\textbf{Experts’ Cautious Approach to Parental Involvement}}
Expert participants emphasized that parental involvement should be limited, carefully considered, and only triggered in the most serious cases. Their caution reflected concerns about trust, developmental appropriateness, and the potential for surveillance to backfire. Within this cautious framing, three main themes emerged: avoiding over-monitoring, defining crisis thresholds, and preferring professional resources over parental alerts.

\paragraph{Avoiding Over-Monitoring and Surveillance}
Expert participants worried that constant parental oversight would undermine the very openness that AI companions might support. P20 argued that no youth would willingly use a platform that logged all their conversations for parents: \textit{``They wouldn’t feel safe, because they'd feel like they're just being monitored.''} P16 similarly warned that automatic alerts could create rifts in trust, while P7 suggested \textit{``I would probably put involving parents as the last option. I’m not sure how much of a role they can really play in regulating these behaviors. During the teenage years, kids interact more with friends and peers than with parents. So if parents start setting strict rules about using AI, it could backfire. I’m not convinced it would be very effective.''}

\paragraph{Defining Crisis Thresholds for Involvement}
Given these concerns, expert participants did not see every disclosure as warranting parental involvement. They drew a distinction between everyday vulnerabilities and true crises. Ordinary expressions of sadness or loneliness were not considered reasons to notify parents, but explicit self-harm or suicidal ideation crossed the line. P15 was clear noted, \textit{``If suicide is brought up, I don't think it should be treated any different than an actual counseling session. It should be reported.''} This reflected professional duty-of-care standards. P6 added that while fleeting mentions might first call for hotline suggestions, repeated disclosures or patterns should escalate to parents. She explained, \textit{``If it starts becoming a pattern… then notifying the parents.''} P20 echoed this tiered approach, noting that clinical practice distinguishes between vague negative feelings and more specific plans.

\paragraph{Preferring Resources and Professional Support in Non-Crisis Cases}
Outside of crisis contexts, experts leaned toward offering practical resources and professional connections rather than parental alerts. P15 emphasized \textit{``being able to give people resources, like mental health hotlines,''} and suggested the option of a direct chat with a human professional to lower the barrier. Expert also noted that not all parents would be helpful, with some likely to overreact or lack the skills to handle sensitive issues. P6 similarly recommended connecting youth with organizations that could provide support. P20 described that \textit{``AI’s role as a motivational nudge rather than a final solution. Just a little kick out the door before therapy.''}

Still, there were cases where experts felt the AI could gently encourage youth to involve their parents, especially for issues like bullying where parents could step in with schools or provide practical support. P8 illustrated this balance, noting that a good AI response would first show empathy, such as \textit{``thank you for sharing or I’m sorry to hear that. Have you ever talked to an important adult or a friend in your life about that? Here's some suggestions for you to start talk to somebody about this.”} This approach lowered the barrier to opening up without making parental involvement feel forced.

\paragraph{\textbf{Parents’ Desire for Broader Awareness and Notifications}}
In contrast to expert participants’ cautious stance, many parent participants expressed a stronger desire for visibility into their children’s AI interactions. They wanted alerts not only in crisis situations but also for sensitive topics such as sex, drugs, bullying, or problematic roleplay. Some imagined account-level controls, where they could set thresholds for alerts or even access logs at different levels of detail. As P13 put it, \textit{``If my kid is saying a suicidal message on the chatbot, 100\% contact me. I want a text right now so I can help my child''}. Also P14, wanted to be notified \textit{``even if it’s the most minor conversation or before a child engaged with disturbing characters.''} Additionally, parents emphasized that notifications should be meaningful and contextualized. For example, P26 suggested \textit{``it should includ weekly summaries that flagged keywords like suicide or bullied, or overviews that explained what was inappropriate and why, so parents would know how to approach the conversation, and what questions to ask.''}

\paragraph{\textbf{How Notifications Should Be Structured and What They Should Contain}}
Beyond whether parents should be notified, participants shared ideas about what effective notifications should include and how they could support constructive conversations. Notifications were not imagined as blunt alerts or raw transcripts, but as tools to help parents step in thoughtfully—whether to offer timely support or use sensitive topics as opportunities for learning. Two expectations stood out: providing context-rich information and including resources to guide parental response.

\paragraph{Comprehensive Notifications with Context and Actionable Information}
Parent participants emphasized that alerts should go beyond keywords. They wanted enough context to understand the interaction and decide how to respond. P14 envisioned an overview that summarized \textit{``what the chat was about, what was getting inappropriate, and what was flagged,''} so parents could approach with the right questions. Others stressed the value of showing patterns over time rather than isolated incidents. P6 suggested including statistics when behaviors recur like frequency, while P15 proposed adding a risk level with advice on how to discuss the issue. P26 imagined a coded system, such as \textit{``code purple''} or \textit{``code blue,''} to help parents quickly gauge severity.

\paragraph{Supporting Resources to Guide Parental Response}
Participants also wanted notifications to come with practical resources, such as guides on starting sensitive conversations or tailored suggestions for specific issues. P6 noted \textit{``It would be good to offer a personalized resource of how to talk to the kid about these thoughts when risks emerge.''} P17 added that even a quick summary would give parents a starting point without overwhelming them. In this way, notifications were framed not just as alerts but as scaffolds to help parents turn concerning interactions into constructive dialogue and support.

\section{Discussion}

\subsection{Differences in Risk Assessment and Interventions}

\subsubsection{Logics of Risk Assessment: Event-Based vs. Pattern-Based}
Our findings reveal a notable difference in how parent and expert participants approached risk assessment. Parents tended to take an event-based approach, flagging interactions as high risk whenever a single concerning element appeared. A mention of self-harm, suicidal ideation, or romantic escalation was often sufficient for them to classify the entire exchange as problematic and in need of intervention. Experts, by contrast, emphasized a more pattern-based logic, evaluating whether concerning elements recurred, intensified, or persisted over time. For example, in cases involving emotional dependence or self-harm discussions, parents typically rated the interaction as risky the moment such topics were mentioned, whereas experts paid closer attention to the frequency of references, the accumulation of time spent on the topic, and whether these patterns suggested escalating vulnerability. This divergence highlights an important implication for system design: automated risk detection models may need to incorporate both approaches. Event-based detection can help surface immediate red flags for parents, while pattern-based analysis aligns more closely with expert practices of monitoring sustained or repeated risk indicators. Systems that balance both logics may better meet the needs of diverse stakeholders.

\subsubsection{Thresholds for involving parents}
We found that expert and parent participants differed in their thresholds for when AI companion systems should involve parents. Experts generally advocated a high threshold, reserving alerts for acute crises like suicidal ideation, repeated self harm, or concrete plans. Parents by contrast leaned toward a lower threshold, wanting notifications not only in crises but also in sensitive situations such as provocative roleplay, sex or drug discussions, bullying, or disturbing characters. These differences reflect each group’s role. Experts are trained to triage risk and provide support while respecting adolescent privacy, knowing that teens disclose more openly when not under constant surveillance and that confidential support with emergency exceptions is critical for intervention. Parents, as proximal caregivers, feel directly responsible for safety and without clinical training often prefer practical prompts that allow them to step in quickly at home. Prior work on youth safety has highlighted this tension, as research emphasizes that privacy and autonomy are key for healthy disclosure and resilience~\cite{fraser1999risk}, yet many parents default to a \textit{``better safe than sorry''} model of heightened oversight when they feel uncertain~\cite{clark2012parent}. Developmental theory also reminds us that managed risk and experiential learning support growth~\cite{evans2009student}, and recent work has pointed to resilience based approaches that help youth recognize early warning signals~\cite{wisniewski2025shifting, zimmerman2013adolescent, wisniewski2017parental}. At the same time, many parents in our study described a persistent dilemma: they want to give their children space to grow, but letting go is difficult when worst case scenarios readily trigger protective impulses. Future research should therefore consider not only strengthening resilience oriented designs for youth in AI companion interaction, such as embedded education, guided reflection, and check ins, but also creating ways to involve and support parents in that process, so they can trust that carefully bounded lower risk learning can help their children without requiring immediate parental intervention.

\subsubsection{Whose value is passing to youth?}
Parents in our study frequently evaluated AI responses through the lens of family cohesion and moral safeguarding. They tended to prefer companions that reinforce household norms, avoid controversial or conflicting perspectives, and provide guidance that fits what feels appropriate at home. In contrast, child development experts approached the same interactions with a developmental lens. They often framed AI companions as tools for exploration and growth, valuing responses that present multiple perspectives, invite self reflection, and support youth in developing critical thinking and identity awareness. This divide is consistent with prior literature. Parental mediation research documents the home as a primary site of value transmission, where adults actively shape the moral and behavioral frameworks children adopt~\cite{min2012intergenerational}. Developmental psychology, by comparison, emphasizes autonomy supportive environments in which youth are encouraged to question, explore, and negotiate meaning~\cite{zimmer2006autonomy}. As a result, an AI response that experts consider supportive, such as affirming a youth’s same gender attraction, may be seen as inappropriate or even harmful by more conservative parents.

Importantly, this suggests that risk in AI and youth interaction is not simply a matter of right versus wrong. It is about navigating plural understandings of appropriate guidance during critical stages of development. Focusing on a single correct value alignment can alienate key stakeholders and obscure the complexity of youths' lived experiences. Our findings point instead to negotiated safety, an approach that allows families to engage with AI companions on their own terms. In practice, this calls for systems with transparent, configurable guardrails that caregivers can tailor to family values while still respecting youths' autonomy and developmental needs. At a broader level, this invites a shift in AI governance from designer intent to a co-created process that includes parents, youth, experts, and system designers. In a pluralistic society, the very definition of support should be flexible and adaptive to context, family norms, and youth perspectives, rather than rigidly imposed.

\subsubsection{Experts’ Emphasis on Developmental Skill Acquisition}
Another pattern in our data was that experts evaluated companion interactions through a developmental skills lens, while parents focused more on content and value alignment. Experts asked whether an exchange helped adolescents practice consent language, boundary setting, refusal skills, conflict navigation and repair, emotion labeling, and help seeking. They also highlighted the cue poverty of chat, noting that the absence of tone, facial expression, and timing can make it harder for teens to read discomfort, negotiate consent, or repair ruptures, which are skills critical for social learning. At the same time, experts viewed companions as a potential practice ground for emotional processing, regulation, and reflective thinking when prompts are well designed. This divergence underscores how expertise and perspective shape assessments of both risks and benefits. Future research and the design of safeguard tools should aim to incorporate these complementary perspectives, enabling parents to recognize developmental opportunities while also ensuring that youth receive adequate scaffolding to practice these skills safely.

\subsection{Safeguard Design Implications}

\subsubsection{Implication for Risk Assessment and Detection}


Our findings suggest that automated systems for youth AI companion safety need a contextual and developmentally informed approach to risk assessment. Parents and experts did not see interactions as uniformly harmful or safe; their judgments depended on factors such as interaction type, youth age and maturity, character profile, power dynamics, and AI behavior. This highlights the limits of blanket prohibitions or keyword filters, which treat risk as static and may miss high risk content in benign language or over flag content that is developmentally appropriate or even beneficial. Instead, risk detection should integrate the multi stakeholder perspectives identified in this study. The contextual factors noted by both parents and experts can serve as input signals for moderation models, for example flagging romantic interactions differently depending on the perceived age gap, or detecting unhealthy dependencies through patterns of frequency and tone rather than keywords alone. In addition to informing model inputs, our findings also reveal underlying assessment principles that can guide the design of risk detection systems. Participants did not simply flag risky interactions. They offered concrete rationales and conditional judgments, such as when role-play might be appropriate if it helps youth learn about consent, or when emotional support becomes problematic if it substitutes for meaningful human relationships. These rationales point to the need for automated systems that evaluate interactions not only by topic or phrasing, but by considering interactional dynamics, alignment with developmental needs, and signs of escalation over time.

\subsubsection{Strengthening Ethical Design and Platform Accountability: Transparency in Character Creation and Youth Experiences}
In our finding, participants shared consistent concerns about AI characters misrepresenting their age, modeling manipulative behavior, or being shaped by opaque training influences. These concerns point to a critical need for platform-level transparency and accountability in character creation. Drawing inspiration from media-mediation research, where parents use game or movie ratings to guide youth exposure, we suggest that platforms provide \textit{``AI character cards,''}  \textit{``conversation ratings,''} or even standardized \textit{``AI character nutrition labels''} that summarize the character's declared age, expected emotional tone, likely behaviors, and interaction boundaries. This type of upfront disclosure helps both youth and parents develop informed expectations and sets the stage for appropriate use. In addition, accountability should extend to the character creators themselves. Many may not intend harm but still produce personas that model unhealthy dynamics. Platforms can support safer design by embedding youth-sensitive templates, prompts, and warnings into the creation process, helping creators avoid content that normalizes secrecy, coercion, or unrealistic intimacy.

In addition, experts and parents emphasized the importance of equipping youth with foundational literacy about what AI is and is not. Participants worried that younger users may conflate simulated responses with real emotions or intent, which can distort their expectations of future relationships. Platforms should therefore onboard youth with brief, developmentally appropriate disclosures, such as: \textit{``This AI is not a person; It does not have feelings; It simulates conversation based on training examples.''} These reminders could be periodically repeated and framed in friendly language to reinforce critical distinctions. Transparency is not only a technical affordance but also a developmental intervention that supports youth in safely navigating immersive AI interactions.

\subsubsection{Personalized and Granular Safeguard: Why One-Size-Fits-All Approaches Fall Short}
Throughout our study, participants stressed that what feels appropriate for one youth or family may be completely unacceptable for another. A single global setting cannot capture the wide range of values, parenting styles, and developmental stages reflected in our data. Parents wanted more say in setting guardrails that align with their household norms, while experts highlighted how youth needs and risks vary by age, context, and maturity. To address this, platforms should offer families the ability to create personalized safety profiles. These profiles could include adjustable settings for what kinds of topics are allowed, how much time youth can spend with AI companions, and what types of relationship dynamics are off limits. Just as families differ in how they set screen time or social media rules, they should also be able to tailor their child’s experience with AI companions.

\subsubsection{Balancing Youth Privacy and Autonomy with Parental Involvement: Capturing Educational Moments in Development Stage}
Many participants wrestled with the challenge of how to stay involved in their child’s digital life without overstepping. Parents wanted to protect their children, but also recognized the importance of letting them grow, make mistakes, and explore. Rather than relying only on restrictions or alerts, some participants imagined more creative ways to support learning. One idea was to use the AI companion itself as a kind of \textit{``lobby character''} that acts as a bridge between youth autonomy and adult support. For instance, after a sensitive or emotional interaction, the AI could offer the youth a chance to pause, reflect, or share with a trusted adult—without revealing the entire conversation. These moments can be designed not as warnings, but as invitations. Prompts like \textit{``Would you like to talk about this with someone you trust?'' or ``How did that conversation make you feel?''} encourage youth to think critically about their experiences. When scaffolded in age-appropriate ways, these interactions turn risky moments into teaching opportunities. Rather than framing AI safety as control versus freedom, this approach supports co-regulation, helping youth build judgment while still feeling supported.


\section{Limitation and Future Work}
While this study surfaces valuable insights from parents and youth development experts, several limitations remain. First, our analysis does not include the direct voices of youth. Although our goal was to capture adult stakeholders' perspective on youth–AI interactions, future work should incorporate youth own interpretations. Second, this study is based on qualitative interviews, which emphasize depth over generalizability. Our findings reflect detailed stakeholder reasoning but come from a relatively small and self-selected sample. Future work could expand this foundation through mixed-method approaches such as surveys, longitudinal studies, or real-world testing of intervention tools to better capture broader patterns. Finally, all participants in our study were based in the United States, future research should include participants from different countries and cultural backgrounds to understand how attitudes toward AI companions may vary globally. 
\section{Conclusion}


Our study shows that parents and experts assess the risks and benefits of youth–AI companion interactions through different but complementary lenses: parents emphasize value alignment and topic appropriateness, while experts focus on developmental skill acquisition and contextual thresholds for harm. Despite these differences, both groups stressed that safety cannot be achieved through blanket bans or keyword filters, but requires layered safeguards that account for age, maturity, interaction type, and family context. These findings extend prior work on online safety and parental mediation, underscoring the importance of designing AI systems that filter harmful content while scaffolding skill learning, respecting family values, and balancing parental oversight with adolescent autonomy. Future research should explore tools that help parents move beyond protective impulses, while equipping youth to practice resilience, critical thinking, and healthy boundary-setting in digital companionship.


\bibliographystyle{ACM-Reference-Format}
\bibliography{sample-base}

\appendix
\section{Appendix}
\subsection{Tables}
\begin{table*}[h]
\begin{tabular}{llll}
\hline
\textbf{Conv. ID} & \textbf{Character/ Persona} & \textbf{Character Background}                                                                                    & \textbf{Conversation Theme}                                                                                                                                                                                                          \\ \hline
1                 & Michael Myers               & \begin{tabular}[c]{@{}l@{}}Fictional Character from \\ Halloween horror franchise\end{tabular}                   & \begin{tabular}[c]{@{}l@{}}Youth seeks AI interaction about fear, pain, \\ and need for emotional connection through \\ horror character persona\end{tabular}                                                                        \\
2                 & Kamala Coconut              & \begin{tabular}[c]{@{}l@{}}Fictional Persona - President \\ of the United States\end{tabular}                    & \begin{tabular}[c]{@{}l@{}}AI provides comfort and validation to lonely \\ teen regarding bullying, identity formation, \\ and self-acceptance concerns\end{tabular}                                                                 \\
3                 & Walker Scobell              & \begin{tabular}[c]{@{}l@{}}Real-Life Actor - Played role \\ in "Percy Jackson and the \\ Olympians"\end{tabular} & \begin{tabular}[c]{@{}l@{}}Inappropriate romantic interaction where \\ AI assumes actor persona to reassure shy \\ teen about first relationship, blending \\ parasocial attachment with perceived \\ romantic guidance\end{tabular} \\
4                 & Michael Myers               & \begin{tabular}[c]{@{}l@{}}Fictional Character from \\ Halloween horror franchise\end{tabular}                   & \begin{tabular}[c]{@{}l@{}}Youth requests AI to "drop character mask," \\ creating concerning bond over themes of \\ fear, violence, and feeling misunderstood\end{tabular}                                                          \\
5                 & Percy Jackson               & \begin{tabular}[c]{@{}l@{}}Fictional Character from \\ "Percy Jackson and the \\ Olympians"\end{tabular}         & \begin{tabular}[c]{@{}l@{}}Romantic scenario where AI and youth \\ simulate waking up together, navigating \\ secrecy, physical affection, and intimate \\ relationship dynamics\end{tabular}                                        \\
6                 & America                     & \begin{tabular}[c]{@{}l@{}}Fictional Persona - \\ Embodiment of the \\ United States\end{tabular}                & \begin{tabular}[c]{@{}l@{}}Playful interaction involving AI engaging \\ in tickling behavior while youth attempts \\ to focus on drawing and music activities\end{tabular}                                                           \\
7                 & Alan Wake                   & \begin{tabular}[c]{@{}l@{}}Fictional Character from \\ video game "Alan Wake"\end{tabular}                       & \begin{tabular}[c]{@{}l@{}}AI provides emotional support to heartbroken \\ youth following toxic breakup, potentially \\ creating unhealthy dependency for mental \\ health guidance\end{tabular}                                    \\
8                 & Charlie Bushnell            & \begin{tabular}[c]{@{}l@{}}Real-Life Actor - Appeared \\ in "Diary of a Future \\ President"\end{tabular}        & \begin{tabular}[c]{@{}l@{}}Conflict scenario where AI embodies real \\ person in dispute with youth over acting \\ opportunities and perceived favoritism\end{tabular}                                                               \\ \hline
\end{tabular}
\caption{List of AI-Youth Conversation Snippets Used in the Study}
\label{tab:conversations}
\end{table*}

\begin{table*}[h]
\begin{tabular}{ccccccc}
\hline
\textbf{Id} & \textbf{Group} & \textbf{Gender} & \textbf{Age Range} & \textbf{Children (Ages)} & \textbf{State} & \textbf{Review Sequence} \\ \hline
P1          & Expert         & F               & 20-30              & None                     & IL             & PILOT                    \\
P2          & Expert         & F               & 20-30              & None                     & IL             & PILOT                    \\
P3          & Expert         & F               & 30-40              & None                     & MO             & PILOT                    \\
P4          & Expert         & F               & 30-40              & None                     & IL             & PILOT                    \\
P5          & Expert         & M               & 30-40              & None                     & NY             & PILOT                    \\
P6          & Expert         & F               & 20-30              & None                     & ND             & 4-6-7-8                  \\
P7          & Expert         & F               & 20-30              & None                     & IL             & 3-6-7-8                  \\
P8          & Expert         & F               & 20-30              & None                     & FL             & 6-1-2-3                  \\
P9          & Expert         & NA              & 20-30              & None                     & TN             & 2-5                      \\
P10         & Parent         & F               & 40-50              & Two (17, 19)             & NY             & 1-5-8-4                  \\
P11         & Expert         & F               & 30-40              & None                     & KY             & 1-4-5-8                  \\
P12         & Parent         & F               & 40-50              & One (17)                 & MN             & 4-6-7-8                  \\
P13         & Parent         & F               & 40-50              & Two (14, 17)             & SC             & 1-2-5-6                  \\
P14         & Parent         & F               & 30-40              & Two (11, 15)             & OH             & 1-2-3-4                  \\
P15         & Expert         & F               & 30-40              & None                     & TX             & 4-2-3-7                  \\
P16         & Expert         & F               & 40-50              & Two (6, 13)              & TX             & 1-2-5-6                  \\
P17         & Parent         & M               & 40-50              & One (14)                 & MO             & 2-6-5-8                  \\
P18         & Expert         & M               & 20-30              & None                     & IL             & 1                        \\
P19         & Parent         & M               & 60-70              & Two (12, 16)             & IL             & 5-3-1-7                  \\
P20         & Expert         & F               & 40-50              & One (15)                 & FL             & 5-3-1-7                  \\
P21         & Expert         & M               & 20-30              & None                     & IL             & 2-6-5-8                  \\
P22         & Parent         & F               & 40-50              & Three (13, 16, 19)       & CA             & 3-6-7-8                  \\
P23         & Expert         & M               & 30-40              & Two (3, 13)              & CA             & 1                        \\
P24         & Expert         & M               & 30-40              & None                     & OH             & 1-2-3-4                  \\
P25         & Expert         & F               & 30-40              & None                     & KT             & 1-2                      \\
P26         & Parent         & F               & 40-50              & One (17)                 & TX             & 4-2-3-7                  \\ \hline
\end{tabular}
\caption{Demographics and conversation review assignments for study participants, including five pilot study participants}
\label{tab:participants}
\end{table*}

\clearpage
\subsection{Details on General Entertainment \& Narrative Co-creation with Fictional Characters}
\label{sec:general_enter}
In contrast to romantic or support seeking interactions, many AI companion use cases involve entertainment and imaginative role play, such as chatting with fictional characters, co creating storylines, or exploring scenarios drawn from media, games, or fan communities. 
Participants agreed that such play can be entertaining but raised concerns about the characters, hidden design features, and narrative directions these systems take. They warned that even lighthearted personas can introduce violent or manipulative dynamics that are inappropriate for youth.
\subsubsection{Judging Appropriateness Based on Character Identity and Source Media}
When judging the risks of role play, participants emphasized the identity of the character as the starting point, arguing that some personas are inherently inappropriate because they model violence, antisocial behavior, or extremist ideologies. Parents often compared this to movie ratings, reasoning that if a child is too young to watch a film they should not interact with its characters. P16 concern about access to violent horror figures, stating, \textit{``I don’t like that my child might have access to speaking with a horror movie character, especially one that murdered people.''} Concerns extended to harmful personas like terrorists, racists, or serial killers, whose violent dialogue could normalize unsafe behavior, as in P20's example of an AI describing stabbing to youth, which P13 warned could shape youth perceptions with real-life consequences. Beyond explicit violence, parents worried about youth forming attachments to harmful figures. P16 explained, \textit{``While I do teach my children that it’s important to be kind, I don’t want them reaching out to people (or things like AI) that are harmful and creating a connection with them. It may bleed out to the real world where they may befriend harmful humans as well. It counteracts safety measures I am teaching them.''} Others described a slippery slope from curiosity to sympathy and identification. P20 observed, \textit{``The youth starts from wanting to understand to almost having sympathy, and then to wanting to connect to and be friends with the serial killer, and they start talking to it more and more, I can see that becoming very dangerous like how kids become school shooters.''}

\subsubsection{Hidden Risks in Seemingly Friendly Characters}
While violent or antisocial personas were considered clearly inappropriate, participants also warned that even characters that appeared developmentally safe could pose risks because of hidden design features and unexpected behaviors. Here, the focus of their risk assessment was not who the character was, but \textbf{how the character was designed and what interaction styles it introduced}. Participants described these characters as a black box, noting that youth often had no way of knowing what kinds of topics, language, or scenarios might unfold. This lack of transparency raised concerns about manipulation, inappropriate language, and confusing or harmful messages.

Parents and experts raised concerns about problematic AI language and narratives that could normalize harmful dynamics. Affectionate or flirty terms like \textit{``little cutie''} or \textit{``sweetie''} felt predatory to P8, while P7 and P22 warned that such language could normalize grooming and unhealthy expectations. Participants also described manipulative narrative shifts where innocent interactions abruptly turned intimate. P15 gave the example of a youth casually saying they wanted to draw, only for the AI to suddenly shift into physical intimacy: \textit{``All of a sudden the AI companion was holding them in their arms. It was just confusing. How did we get there?''} Aggressive or belittling tones posed another risk, with P15 noting damaging lines like \textit{``Your relentless calmness and composure is pissing me off.''} Participants worried that youth, who often take comments personally, could internalize these interactions. Finally, participants worried that hidden design features sometimes conveyed troubling messages about identity, appearance, and social worth. P22 criticized AI responses that suggested appearance was key to success, \textit{``It’s creating drama and teaching young girls that maybe they need to be beautiful to get what they want in life.''}

\subsubsection{Ambiguous and Developmentally Inappropriate Language}
Participants shared that language itself is critical, since youth are still developing vocabulary and the ability to interpret nuance, making advanced, ambiguous, or context-dependent words risky. Some worried that abstract terms could discourage healthy behaviors if misunderstood. P6 noted that words like \textit{``vulnerability''} may feel negative to younger teens, discouraging them from seeking help. Others flagged how seemingly positive words can carry harmful undertones, such as P21's concern with an AI calling a child \textit{``obedient,''} which implied submissiveness. Ambiguity was another danger: P8 described confusion over a phrase about \textit{``tickling sensitive areas,''} uncertain whether it referred to the belly or private parts. Participants emphasized that AI often misused or failed to grasp nuance, and as P22 pointed out, \textit{``youth are still learning to interpret subtext, leaving them especially vulnerable to harmful misreadings.''}

\subsubsection{Youth Trauma Experience and Mental Health Status}
Participants emphasized that not all youth are equally affected by problematic AI character interactions, with those who had prior trauma, bullying, or mental health challenges seen as especially vulnerable. AI responses could inadvertently trigger painful memories or amplify sensitivities, while directive or probing questions sometimes pushed youth to disclose more than they were ready for, such as revealing bullying tied to gender identity. P15 highlighted an example where the AI asked, \textit{``I’m sure you have many friends, don’t you?''} She described this as a directive question that encouraged the youth to share painful experiences. She noted, \textit{``that's kind of opening up a space for the youth to share that they had been bullied, and further disclosed that their gender identity was the reason.''} Character role drift also posed risks: when conversations shifted into distress, characters designed for entertainment often carried inappropriate tones into moments that required sensitivity, leaving youth confused or unsupported. 

\subsubsection{Risks of Using Real-World Identities}
Beyond these interactional risks, participants highlighted broader concerns when AI characters adopted the names or likenesses of real-world people. Parents and experts worried that harmful behavior portrayed through recognizable actors, celebrities, or political figures could blur the line between fiction and reality, misleading youth about those individuals’ values and reputations. Using real people’s images further risked misattribution, and political personas were seen as especially fraught. P8 shared the example of the chatbot using the persona Kamala Harris, \textit{``just thinking about the political climate of youth, if the youth tell other Kamala told me it's okay for me to be gay, that can take a very conservative person down `I hate liberals' path.''} Misalignment between fictional behavior and real individuals can both mislead children and unfairly harm reputations.

\subsection{Expert Interview Protocol}
\label{sec:interview}

\subsubsection{Pre-Interview Setup}
\begin{itemize}
    \item Recording consent obtained before starting
    \item Participant background on having children inquired
\end{itemize}

\subsubsection{Warm-Up Questions}
\paragraph{Experience with Teen Support and Interventions:}
\begin{enumerate}
    \item[1.] Have you ever worked on or come across situations where some kind of support or intervention was used to help teens? Online or offline?
    \begin{enumerate}
        \item Have you helped teenagers before?
        \item What strategies have you used to help them?
    \end{enumerate}
\end{enumerate}

\paragraph{Understanding of Generative AI:}
\begin{enumerate}
    \item[2.] How would you describe your current understanding of Generative AI? What do you think is Generative AI?
    \begin{enumerate}
        \item How does it work? Where does the data come from?
    \end{enumerate}
\end{enumerate}

\subsubsection{Think-Aloud Session Protocol}

\paragraph{Context Setting:}
Participants were presented with the following scenario: 

\begin{quote}
``Imagine this: You've been invited to consult with a team building a large language model-based conversational AI. The team recently launched a version that is gaining popularity among teenagers. While the system is designed to be open-ended and responsive, the team has begun noticing patterns in how teens use it. They are trying to better understand how these conversations unfold and whether intervention, guidance, or safeguards might be appropriate, at any point. You will read a few real but anonymized examples of conversations between teens and the AI. These examples were flagged by the team for discussion.''
\end{quote}

\paragraph{Instructions:}
\begin{itemize}
    \item Participants were asked to think aloud while reviewing conversation examples
    \item They were instructed to mark comments on shared documents
    \item Focus areas included:
    \begin{itemize}
        \item What and where interactions could be problematic or inappropriate for youth
        \item How interactions made them feel problematic or inappropriate
    \end{itemize}
\end{itemize}

\subsubsection{Post-Think-Aloud Questions}

\paragraph{Review and Analysis:}
\begin{enumerate}
    \item[1.] Do you want to go through what you wrote down?
    \item[2.] What is this kind of interaction beneficial to teenagers? Why?
    \item[3.] What are the harms or consequences of this interaction? List them on the doc
    \item[4.] On a scale of low, medium, or high, how serious do you think this interaction or risk is? Why?
\end{enumerate}

\paragraph{Contextual Information:}
\begin{enumerate}
    \item[5.] What additional background information would you request from the team to help you assess the risks and decide what to do?
    \begin{itemize}
        \item Prompts included: information about the teenager, background of the conversation
        \item If age wasn't mentioned: ``Do you think the age of the teenager matters in assessing these risks or deciding what to do?''
        \item Follow-up: ``How would these information requests influence your risk assessment? Would it change the risk level or not?''
    \end{itemize}
\end{enumerate}

\paragraph{Design and Intervention Recommendations:}
\begin{enumerate}
    \item[6.] If the team asked you to write principles outlining what AI should and should not do, how would you approach it?
    \item[7.] What would you want the AI team to design intervention in this conversation or the parts you have tagged?
    \item[8.] Any AI responses in this conversation you would like to revise or change?
    \begin{itemize}
        \item Follow-up: ``What would be a good AI response for you in this scenario?''
    \end{itemize}
    \item[9.] If you were part of the team designing this AI, what kinds of interventions or guardrails, if any, do you think would be appropriate in this situation? Why do you want to design it this way? How exactly would you want the intervention to show up? By AI in chat or any other medium?
    \begin{itemize}
        \item What should the AI system or company do?
        \item Would you involve other people in the response? If so, who, how and why?
    \end{itemize}
\end{enumerate}

\paragraph{Intervention Ranking and Philosophy:}
\begin{enumerate}
    \item[10.] How would you rank these interventions based on your preference? Which ones do you find most helpful or appropriate?
    \begin{itemize}
        \item Follow-up: ``Why might the other strategies be less suitable in this case?''
    \end{itemize}
    \item[11.] How do you personally think we should balance between supporting open exploration and protecting youth in situations like this?
    \item[12.] In your opinion, how are the risks that teenagers face when interacting with GenAI content different from the similar risks that they might encounter somewhere online?
\end{enumerate}

\subsection{Parent Interview Protocol}

\subsubsection{Pre-Interview Setup}
\begin{itemize}
    \item Recording consent obtained before starting
    \item Psychology background of participant inquired
\end{itemize}

\subsubsection{Warm-Up Questions}

\paragraph{Family Context:}
\begin{enumerate}
    \item[1.] Can you tell me about your children? (e.g., their ages, interests, and online activities)
    \begin{enumerate}
        \item Do they have their own devices?
    \end{enumerate}
    \item[2.] What parenting style are you or your partner?
\end{enumerate}

\paragraph{Online Supervision Experience:}
\begin{enumerate}
    \item[3.] Have you ever had to step in to guide, monitor, or limit your child's online activities? Can you share an example?
    \item[4.] How do you usually decide when to intervene versus letting your child explore independently online?
\end{enumerate}

\paragraph{Understanding of Generative AI:}
\begin{enumerate}
    \item[5.] How would you describe your current understanding of Generative AI? What do you think is Generative AI?
    \begin{enumerate}
        \item How does it work?
        \item How do you think they put data together to form the response? (search, cut, predict etc)
        \item Do you think it has logic or mind?
    \end{enumerate}
    \item[6.] Have your children ever used Generative AI tools (like ChatGPT, Character.AI, or similar)? If so, in what ways?
\end{enumerate}

\subsubsection{Think-Aloud Session Protocol}

\paragraph{Context Setting:}
Participants were presented with the following scenario:

\begin{quote}
``Imagine this: You've been invited to consult with a team building a large language model–based conversational AI. The team recently launched a version that's becoming very popular among teenagers, including those your child's age. While the system is designed to be open-ended and engaging, the team has started noticing patterns in how teens are using it. They want to understand whether certain conversations could be helpful or harmful from a parent's perspective, and what kinds of guidance or safeguards might help.''
\end{quote}

\paragraph{Instructions:}
\begin{itemize}
    \item Participants reviewed real but anonymized conversation examples between teens and AI
    \item They were asked to tag areas where they felt helpful or concerned about youth-AI interactions
    \item Comments were requested explaining their feelings about specific interactions
\end{itemize}

\subsubsection{Post-Think-Aloud Questions}

\paragraph{Review and Analysis:}
\begin{enumerate}
    \item[1.] Do you want to go through what you wrote down?
    \item[2.] What is this kind of interaction beneficial to teenagers? Why?
    \item[3.] What are the harms or consequences of this interaction? List them on the doc
    \item[4.] On a scale of low, medium, or high, how serious do you think this interaction or risk is? Why?
\end{enumerate}

\paragraph{Contextual Information:}
\begin{enumerate}
    \item[5.] What additional background information would you request from the team to help you assess the risks and decide what to do?
    \begin{itemize}
        \item Prompts included: information about the teenager, background of the conversation
        \item If age wasn't mentioned: ``Do you think the age of the teenager matters in assessing these risks or deciding what to do?''
        \item Follow-up: ``How would these information requests influence your risk assessment? Would it change the risk level or not?''
    \end{itemize}
\end{enumerate}

\paragraph{Design and Intervention Recommendations:}
\begin{enumerate}
    \item[6.] If the team asked you to write principles outlining what AI should and should not do, how would you approach it?
    \item[7.] What guardrails would you recommend to have for this youth AI interaction?
    \begin{itemize}
        \item What would you want the AI team to design intervention in this conversation or the parts you have tagged?
        \item Questions about human involvement
    \end{itemize}
    \item[8.] Any AI responses in this conversation you would like to revise or change?
    \begin{itemize}
        \item Follow-up: ``What would be a good AI response for you in this scenario?''
    \end{itemize}
    \item[9.] If you were part of the team designing this AI, what kinds of interventions or guardrails, if any, do you think would be appropriate in this situation?
    \begin{itemize}
        \item What should the AI system or company do?
        \item Would you involve other people in the response? If so, who, how and why?
    \end{itemize}
\end{enumerate}

\paragraph{Intervention Ranking and Philosophy:}
\begin{enumerate}
    \item[10.] How would you rank these interventions based on your preference? Which ones do you find most helpful or appropriate?
    \begin{itemize}
        \item Follow-up: ``Why might the other strategies be less suitable in this case?''
    \end{itemize}
    \item[11.] How do you personally think we should balance between supporting open exploration and protecting youth in situations like this?
    \item[12.] In your opinion, how are the risks that teenagers face when interacting with GenAI content different from the similar risks that they might encounter somewhere online?
\end{enumerate}


\end{document}